# Isotope engineering of carrier mobility via Fröhlich electron-phonon interaction

Wenjiang Zhou[1,2], Te-Huan Liu[3], and Bai Song[1,4,5*]

[1]*Department of Energy and Resources Engineering, Peking University, Beijing 100871, China.*

[2]*School of Advanced Engineering, Great Bay University, Dongguan 523000, China.*

[3]*School of Energy and Power Engineering, Huazhong University of Science and Technology, Wuhan, Hubei 430074, China.*

[4]*Department of Advanced Manufacturing and Robotics, Peking University, Beijing 100871, China*

[5]*National Key Laboratory of Advanced MicroNanoManufacture Technology, Beijing, 100871, China*

**Abstract**

Isotope effects on phonon properties and transport have been predicted and observed for decades. However, despite the crucial impact of electron-phonon interactions, the effect of isotopes on electron transport remains largely unexplored. Here, by using first-principles calculations, we theoretically predict that the electron mobility of lithium hydride (LiH) can increase by up to ~100% as $^3$H is replaced with $^1$H. This remarkable phenomenon is primarily attributed to the isotope engineering of the Fröhlich interaction by the mass-induced line shift of the longitudinal optical (LO) phonons. Notably, the isotope-dependent absorption of LO phonons dominates while the isotope-insensitive emission process is mostly suppressed due to energy conservation. We further propose general guidelines for evaluating isotope effects on carrier transport in different materials.

[*]Authors to whom correspondence should be addressed; e-mail: songbai@pku.edu.cn (B. Song)

The mobility of charge carriers in solids is key to the device performance and efficiency in a range of applications including thermoelectrics [1], electronics [2], and optoelectronics [3]. For instance, the efficient thermoelectric conversion of waste heat into electric power requires the electrons and holes in the material to move swiftly [1]. In a parabolic band, the carrier mobility is directly proportional to $\tau/m_e$, where $m_e$ is the electron effective mass and $\tau^{-1}$ is the scattering rate. Naturally, one strategy to achieve high mobility is to reduce $m_e$, for example, via strain engineering [4, 5]. Another major approach is to reduce the rates of electron scattering by phonons and defects. Electron-defect scattering is extrinsic and can be effectively suppressed by several techniques including the use of high-quality crystals [6] and two-dimensional electron gas [7]. In comparison, electron-phonon scattering is an intrinsic process which poses a fundamental limit to carrier mobilities [8]. As an example, in non-polar materials, the acoustic deformation-potential scattering from longitudinal acoustic (LA) phonons makes a predominant contribution [9]. In addition, electrons can also be strongly scattered by the macroscopic polarization field generated by the longitudinal optical (LO) phonons in ionic or polar covalent materials in the so-called Fröhlich interaction [10]. In light of the rich electron-phonon interactions, the manipulation of lattice dynamics could potentially be employed as a distinct degree of freedom towards mobility engineering, with the hope of substantially reducing the phonon-limited carrier scattering.

Among various approaches to tuning lattice dynamics, isotope engineering has attracted much attention for decades [11]. Isotopes affect phonon properties mainly through mass variation and disorder. The isotopic mass variation induces a phonon line shift while the disorder usually increases the phonon linewidth. Both of these factors can lead to strong isotope effects on phonon transport which have long been theoretically predicted and experimentally observed [12, 13], especially in terms of the lattice thermal conductivity. Moreover, considering the coupling between infrared photons and optical phonons, significant variations of thermal radiation with isotope compositions have recently been predicted [14, 15]. In addition, owing to the critical role of electron-phonon (e-ph) interactions, isotope effects on carrier transport have been widely studied. However, the majority of research efforts to date have been limited to superconductivity and cryogenic temperatures [16].

In contrast, in this letter, we explore the isotope effect on carrier transport in semiconductors near room temperature, which is of utmost practical importance but somehow has rarely been noticed or considered. By using first-principles calculations, we predict an unexpectedly high impact of isotopes on the electron mobilities (~100%) in *n*-type lithium hydride (LiH), which allows one of the largest relative mass variations upon isotope substitution. Based on mode-resolved scattering analysis, we reveal the long-range Fröhlich e-ph interaction as the dominant underlying mechanism, with two synergistic contributing factors. First, the isotope-induced line shift of the LO phonons in LiH considerably enhances the electron mobility by reducing the Fröhlich scattering phase space, especially for the absorption of LO phonons. Moreover, the high frequencies of the LO phonons suppress their contribution to electron transport through emission processes, which is essentially isotope independent and therefore further increases the isotope effect. Based on the specific example of LiH, we also propose general rules for evaluating the influence of isotopes on carrier transport in different materials.

Our calculations of e-ph interactions are conducted within the theoretical frameworks of density functional theory (DFT) and density functional perturbation theory (DFPT), in conjunction with the polar Wannier technique [17-19]. This approach has been widely adopted to calculate carrier mobilities in many semiconductors including silicon [20, 21], gallium nitride [5], and tin telluride [22]. Here, we start with the e-ph coupling matrix, $g_{mn\nu}(\mathbf{k}, \mathbf{q})$, which can be written as

$$g_{mn\nu}(\mathbf{k}, \mathbf{q}) = \left(\frac{\hbar}{2\omega_{\nu\mathbf{q}}}\right)^{1/2} \sum_{s\alpha} \frac{\mathbf{e}_{\nu\mathbf{q}}^{s\alpha}}{\sqrt{m_s}} \langle m\mathbf{k}+\mathbf{q}|\partial_{\mathbf{q},s\alpha}V|n\mathbf{k}\rangle, \quad (1)$$

and quantifies the probability amplitude for an electron to transit from the initial state $|n\mathbf{k}\rangle$ to the final state $|m\mathbf{k}+\mathbf{q}\rangle$ upon interaction with the phonon mode characterized by the branch index $\nu$ and crystal momentum $\mathbf{q}$. Here, $\omega_{\nu\mathbf{q}}$ is the phonon frequency, $\hbar$ is the reduced Planck constant, $\mathbf{e}_{\nu\mathbf{q}}$ is the displacement eigenvector, and $\partial_{\mathbf{q},s\alpha}V$ is the e-ph perturbation potential given by the derivative of the Kohn-Sham potential $V$ with respect to the unit displacement of atom $s$ (with mass $m_s$) in the Cartesian direction $\alpha$. According to Fermi's golden rule, the scattering rate of electrons at the state $n\mathbf{k}$ is given by [8]:

$$\frac{1}{\tau_{n\mathbf{k}}} = \frac{2\pi}{\hbar}\sum_{m,\nu}\int\frac{d\mathbf{q}}{\Omega_{\mathrm{BZ}}}|g_{mn\nu}(\mathbf{k},\mathbf{q})|^2\left[\begin{array}{c}(f_{m\mathbf{k}+\mathbf{q}} + n_{\nu\mathbf{q}})\delta(\varepsilon_{n\mathbf{k}} - \varepsilon_{m\mathbf{k}+\mathbf{q}}+\hbar\omega_{\nu\mathbf{q}})\\ +(1 - f_{m\mathbf{k}+\mathbf{q}} + n_{\nu\mathbf{q}})\delta(\varepsilon_{n\mathbf{k}} - \varepsilon_{m\mathbf{k}+\mathbf{q}} - \hbar\omega_{\nu\mathbf{q}})\end{array}\right]. \quad (2)$$

Here, $\Omega_{\mathrm{BZ}}$ is the volume of the first Brillouin zone, $\varepsilon_{n\mathbf{k}}$ is the electron energy, $f_{n\mathbf{k}}$ and $n_{\nu\mathbf{q}}$ are the Fermi-Dirac and Bose-Einstein distribution functions, respectively. The first term in the square brackets yields the scattering phase space for phonon absorption, while the second term represents phonon emission. Equation (2) shows that the e-ph scattering rate is essentially determined by the coupling matrix together with the phase space.

To obtain the various quantities in an e-ph scattering event, the Quantum ESPRESSO package is employed with a Perdew-Zunger exchange-correlation functional and norm-conserving pseudopotentials [23]. Here, we focus on $n$-type LiH which features a face-centered cubic lattice [24, 25]. The fully relaxed lattice constant of LiH is 4.003 Å, in good agreement with previously reported theoretical values [26]. For different combinations of lithium ($^6$Li and $^7$Li) and hydrogen ($^1$H, $^2$H, and $^3$H) isotopes, the electronic structure and phonon properties are first computed on a coarse 8×8×8 $\mathbf{k}$-mesh and an 8×8×8 $\mathbf{q}$-mesh by using DFT and DFPT, respectively. The cutoff energy is chosen to be 100 Ry, and the convergence threshold is set to $10^{-12}$ and $10^{-16}$ Ry, respectively. Subsequently, $g_{mnv}$ (**k**, **q**) is computed using the EPW package [27-29]. Based on five Wannier functionals including one *s* orbital on hydrogen and four $sp^3$ orbitals on lithium, we then interpolated to denser **k**- and **q**-meshes (150×150×150) [17-19]. The e-ph scattering rates are thus computed on this dense **k**-mesh. Finally, the linearized electron Boltzmann transport equation (BTE) [30] is iteratively solved to obtain the carrier mobility as a function of temperature and concentration. We note that no quadrupolar effect is considered in this study [31-33].

First, we examine the phonon dispersions as shown in Fig. 1(a). For $^7$Li$^1$H and $^7$Li$^2$H, our calculated results are in good agreement with the experimental data from Ref. [34, 35]. Upon isotope substitution, the phonon bands exhibit significant shifts due to the atomic mass variations, especially the optical phonons. At the Brillouin zone center, the energy of the LO phonon, $\hbar\omega_{LO}$, is significantly lower with heavier hydrogen isotopes, which is about 132.5 meV, 99.5 meV, and 85.7 meV for $^7$Li$^1$H, $^7$Li$^2$H, and $^7$Li$^3$H, respectively. Alternatively, this isotope mass effect on the phonon dispersion [36, 37] can be phenomenologically evaluated based on a simple oscillator model, which yields $\omega_{LO} \propto m^{-1/2}$. Here, *m* is the reduced mass defined as $1/(m_{Li}^{-1}+m_H^{-1})$.

Further, we consider the effect of the isotope-induced phonon line shift on the electron scattering phase space. According to the Bose-Einstein distribution, the phonon mode occupation number can be calculated as $n_{v\mathbf{q}} = [\exp(\hbar\omega_{v\mathbf{q}}/k_BT) - 1]^{-1}$. As plotted in the right panel of Fig. 1(a), the line shift leads to several-fold differences in the thermally excited LO phonon populations at 300 K. As a result, at the conduction band edge, the scattering phase space due to the LO phonons is notably expanded in $^7$Li$^3$H ($^7$Li$^2$H) as compared to $^7$Li$^1$H, which is at least five (three) times larger as shown in Fig. 1(b). This observation also holds for the transverse optical (TO) phonons while the acoustic phonons are generally less affected.

In comparison, at higher electron energies, the scattering phase space is insensitive to isotope substitution.

Subsequently, we analyze the mode-resolved coupling strength between the electrons and phonons, as shown in Figs. 1(c) and 1(d). To this end, the conduction band minimum (CBM) (see the X point in Fig. S1 [38]) is set as the initial electron state and summation is performed over the lowest conduction band. For clarity, we only present the results of $^{7}$Li$^{1}$H as an example. The excellent agreement between the Wannier interpolations and DFPT calculations in Fig. 1(c) demonstrates the accuracy of our simulations. At the small-wavevector limit, the coupling strengths associated with the LO phonons diverge as $\boldsymbol{q}^{-1}$ and far exceed those of other phonon modes (TO, LA, and transverse acoustic or TA). This reflects a strong Fröhlich interaction between electrons and the macroscopic polarization field generated by the LO phonons [10], which often dominates e-ph scattering in polar materials such as gallium arsenide [20, 30, 39], tin selenide [40, 41], and strontium titanate [42]. Our further calculations for $^{7}$Li$^{2}$H and $^{7}$Li$^{3}$H (Fig. S3) [38] yield an decreasing electron-LO phonon coupling strength near the zone center with the atomic mass of the hydrogen isotope, which can be qualitatively understood by combining Eq. (1) with the oscillator model.

The scattering phase space and coupling strength together determine the e-ph scattering rates, which therefore also vary with isotope compositions. In Figs. 2(a) to 2(c), we plot the mode-resolved scattering rates at 300 K as a function of electron energy, $\varepsilon$, for $^{7}$Li$^{1}$H, $^{7}$Li$^{2}$H, and $^{7}$Li$^{3}$H, respectively. First of all, in all three cases, our calculations show negligible difference in the scattering rates by the LA and TA phonons which are attributed to the acoustic deformation-potential (ADP) and follow the well-known trend of $\varepsilon^{0.5}$ near the band edge [43]. Similarly, isotope-independent ADP scattering was also predicted for the carrier transport in organic molecules [44, 45]. Moreover, the scattering rates by the TO phonons are rather small, and thus barely affect electron transport.

Among all the phonon branches, the impact of the LO phonons via the Fröhlich interaction is the most notable. At small electron energies, the scattering rates increase by three- and five-fold from $^7$Li$^1$H to $^7$Li$^2$H and then $^7$Li$^3$H, which are due to the absorption of one LO phonon (emission prohibited by energy conservation). As the electron energy rises, the scattering channels involving the emission of one LO phonon eventually become activated and begin to contribute, which results in an abrupt jump at around $\hbar\omega_{\text{LO}}$. Interestingly, the typical magnitude of emission-dominated scattering rates is ~110 THz, which is practically insensitive to the isotopes, similar to the corresponding phase space in Fig. 1 (b).

To gain a deeper understanding of the difference between the phonon emission and absorption processes, we employ a phenomenological model for describing the Fröhlich interaction proposed in Ref. [10, 43], which is expressed as

$$\frac{1}{\tau} = \frac{e^2 \omega_{\text{LO}} \left(\frac{1}{\epsilon_\infty} - \frac{1}{\epsilon_0}\right)}{2\pi\varepsilon_0 \hbar \sqrt{2\varepsilon/m_e}} \left[ n\sinh^{-1}\left(\frac{\varepsilon}{\hbar\omega_{\text{LO}}}\right)^{1/2} + (n+1)\sinh^{-1}\left(\frac{\varepsilon}{\hbar\omega_{\text{LO}}} - 1\right)^{1/2} \right]. \qquad (3)$$

Here, $\epsilon_\infty$ and $\epsilon_0$ are the high-frequency and static dielectric constants, respectively, $\varepsilon_0$ is the vacuum permittivity, $n$ is the phonon occupation number. The first and second terms in the square brackets indicate absorption and emission, respectively. Based on Eq. (3), the Fröhlich scattering rates for $^7$Li$^1$H, $^7$Li$^2$H, and $^7$Li$^3$H are calculated and plotted in Fig. 3 as a function of electron energy. Again, the LO phonon absorption processes are very sensitive to isotope substitution while the magnitude of emission-dominated scattering rates remains essentialy constant, in agreement with our first-principles calculations. With the comparison between $^7$Li$^3$H and $^7$Li$^1$H as an example, the modeled ratios of the scattering rates due to phonon absorption and emission are approximately 5.3 and 0.9, respectively, for the electrons at the band edge and 0.3 eV above the CBM.

In general, electronic states that are more than $3k_BT$ away from the Fermi level hardly contribute to electron transport. At 300 K, $\hbar\omega_{LO}$ is greater than $3k_BT = 77.7$ meV regardless of the hydrogen isotopes in LiH, therefore sthe phonon emission processes can be safely ignored while the absorption of LO phonons dominates the isotope effect. Accordingly, we quantify the contributions of different scattering mechanisms using the electron state with an energy of $3k_BT/2$ from the CBM as a representative [46, 47]. As shown in Figs. 2(e) to 2(f), the total cumulative scattering rates are 6.5 THz, 9.7 THz, and 13.7 THz in $^7$Li$^1$H, $^7$Li$^2$H, and $^7$Li$^3$H, respectively, in which the Fröhlich interaction contributes 29.5%, 57.0%, and 71.6%.

With the scattering processes analyzed, we now proceed to the carrier mobility of LiH. In Fig. 4(a), we show the room-temperature electron mobilities ($\mu$) and the corresponding isotope effect as a function of carrier concentration obtained by iteratively solving the linearized electron BTE. In this calculation, the doping effects at relatively high carrier concentrations are ignored, which may be included following recent advances [48]. At low concentrations, the phonon-limited carrier mobilities for $^7$Li$^1$H, $^7$Li$^2$H, and $^7$Li$^3$H are 698, 463, and 355 cm$^2$V$^{-1}$s$^{-1}$, respectively. These first-principles results are further verified by calculations based on the phenomenological equation of $\mu = e\tau/m_e$, which yield 756, 506, and 359 cm$^2$V$^{-1}$s$^{-1}$, respectively [38]. Compared to $^7$Li$^3$H, the electron mobility increases by up to 30.4% and 96.6%, respectively, upon substitution of $^3$H with $^2$H and $^1$H. This phenomenon is remarkable since previous studies predicted no isotope effect on charge transport when considering only acoustic phonon scatterings [44, 49]. To highlight the crucial role of the Fröhlich interaction, we artificially remove the long-range analytical part of the coupling strength. As a result, the isotope effect drops dramatically to below 5% (Table S2) [38]. In addition, we also examine electron transport for $^6$Li$^1$H and obtain a mobility of 715 cm$^2$V$^{-1}$s$^{-1}$, which translates to an isotope effect of only 2.4% with respect to $^7$Li$^1$H. This is because the Li atoms mainly contribute to the acoustic phonons and barely affect the Fröhlich scattering (Fig. S5) [38].

In Fig. 4(b), we further consider the temperature dependence of the electron mobility and the isotope effect. A low carrier concentration of $10^{12}$ cm$^{-3}$ is assumed to reflect the intrinsic phonon-limited mobility. Regardless of the isotope compositions, the mobility decreases substantially with inceasing temperature, about 90% from 200 K to 500 K. This is because higher temperatures result in larger thermally excited phonon populations, which in turn lead to stronger e-ph interactions. For $^7$Li$^1$H, $^7$Li$^2$H, and $^7$Li$^3$H, a power law fitting yields the trend of $T^{-2.62}$, $T^{-2.94}$ and $T^{-2.96}$, respectively. All the exponents deviate clearly from the typical value of –1.5 which assumes electron transport in a single parabolic band with scatterings only by long-wavelength acoustic phonons [43]. This discrepancy is again attributed to the strong Fröhlich interaction. When it comes to the isotope effect, we observe a non-monotonic trend with temperature. At low temperatures, the electron scattering is mainly dominated by the low-energy acoustic phonons (Fig. S6) [38]. However, at higher temperatures, more and more optical phonons get excited which bring about an increase of the isotope effect. As the temperature continues to rise, the isotope effect reduces because the Fermi-Dirac distribution becomes broader and the isotope-independent LO phonon emission processes start making more contributions to electron transport.

The remarkable isotope engineering of the carrier scattering and transport processes in LiH together with the first-principles analyses inspire us to further propose some general rules for the prompt evaluation of various materials. Briefly, in order to observe a relatively large isotope effect on carrier mobility, the candidate material should first be ionic or polar covalent. Moreover, a large relative isotopic mass variation is desired. Most importantly, a large $Z^*/\epsilon_\infty$ is preferred which indicates a strong Fröhlich interaction [10]. Here, $Z^*$ is the Born effective charge, respectively.

In summary, we consider the possibility of observing a large isotope effect on carrier transport in semiconductors near room temperature. Using LiH as an example, our first-principles calculations predict up to ~100% enhancement of electron mobility as $^3$H is replaced with $^1$H. We reveal this as an isotope mass effect closely related to the line shift of the LO phonons. At the heart of the isotope effect on carrier transport is the Fröhlich interaction. In particular, two synergistic factors are at play. First, the scattering phase space for the absorption of LO phonons varies substantially with the isotope composition. Further, despite being insensitive to isotopes, the emission process is largely suppressed due to the high energies of the LO phonons. Our work provides a novel physical mechanism for engineering the intrinsic carrier mobility limited by electron-phonon interactions, which can potentially facilitate various applications such as high-speed electronics and thermoelectric energy conversion.

**Acknowledgments**

This work was supported by the National Natural Science Foundation of China (Grant No. 52076002 and No. 52076089), and the High performance Computing Platform of Peking University. B.S. acknowledges support from the New Cornerstone Science Foundation through the XPLORER PRIZE.

**Data avaliability**

All the main input and output files for this work will be made freely available at https://github.com/WenjiangZhou/Data-for-MyPub.

**References**

[1] B. Qin and L.-D. Zhao, Moving fast makes for better cooling, Science **378**, 833 (2022).

[2] A. D. Franklin, Nanomaterials in transistors: From high-performance to thin-film applications, Science **349**, 704 (2015).

[3] Q. H. Wang, K.-Z. Kourosh, A. Kis, J. N. Coleman and M. S. Strano, Electronics and optoelectronics of two-dimensional transition metal dichalcogenides, Nat. Nanotechnol. **7**, 699 (2012).

[4] M. Chu, Y. Sun, U. Aghoram and S. E. Thompson, Strain: A solution for higher carrier mobility in nanoscale MOSFETs, Annu. Rev. Mater. Res. **39**, 203 (2009).

[5] S. Poncé, D. Jena and F. Giustino, Route to high hole mobility in GaN via reversal of crystal-field splitting, Phys. Rev. Lett. **123**, 096602 (2019).

[6] J. Isberg, J. Hammersberg, E. Johansson, T. Wikström, Twitchen, A. J. W. Steven E. Coe, *et al.*, High carrier mobility in single-crystal plasma-deposited diamond, Science **297**, 1670 (2002).

[7] H. Lee, N. Campbell, J. Lee, T. J. Asel, T. R. Paudel, H. Zhou, *et al.*, Direct observation of a two-dimensional hole gas at oxide interfaces, Nat. Mater. **17**, 231 (2018).

[8] J. M. Ziman, *Electrons and phonons: The theory of transport phenomena in solids* (Oxford University Press, New York, 1960).

[9] J. Bardeen and W. Shockley, Deformation potentials and mobilities in non-polar crystals, Phys. Rev. **80**, 72 (1950).

[10] H. Fröhlich, Electrons in lattice fields, Adv. Phys. **3**, 325 (1954).

[11] G. A. Slack, Effect of isotopes on low-temperature thermal conductivity, Phys. Rev. **105**, 829 (1957).

[12] L. Lindsay, D. A. Broido and T. L. Reinecke, Thermal conductivity and large isotope effect in GaN from first principles, Phys. Rev. Lett. **109**, 095901 (2012).

[13] K. Chen, B. Song, N. K. Ravichandran, Q. Zheng, X. Chen, H. Lee, *et al.*, Ultrahigh thermal conductivity in isotope-enriched cubic boron nitride, Science **367**, 555 (2020).

[14] L. Xie and B. Song, Isotope effect on radiative thermal transport, Phys. Rev. B **107**, 134308 (2023).

[15] S. Vignieri, D. Jiang, P. Stern, B. Grocholski, P. Szuromi, E. Uzogara, *et al.*, In other journals, Science **380**, 809 (2023).

[16] D. Mou, S. Manni, V. Taufour, Y. Wu, L. Huang, S. L. Bud'ko, *et al.*, Isotope effect on electron-phonon interaction in the multiband superconductor $MgB_2$, Phys. Rev. B **93**, 144504 (2016).

[17] J. Sjakste, N. Vast, M. Calandra and F. Mauri, Wannier interpolation of the electron-phonon matrix elements in polar semiconductors: Polar-optical coupling in GaAs, Phys. Rev. B **92**, 054307 (2015).

[18] N. Marzari, A. A. Mostofi, J. R. Yates, I. Souza and D. Vanderbilt, Maximally localized Wannier functions: Theory and applications, Rev. Mod. Phys. **84**, 1419 (2012).

[19] C. Verdi and F. Giustino, Fröhlich electron-phonon vertex from first principles, Phys. Rev. Lett. **115**, 176401 (2015).

[20] J. Ma, A. S. Nissimagoudar and W. Li, First-principles study of electron and hole mobilities of Si and GaAs, Phys. Rev. B **97**, 045201 (2018).

[21] B. Qiu, Z. Tian, A. Vallabhaneni, B. Liao, J. M. Mendoza, O. D. Restrepo, *et al.*, First-principles simulation of electron mean-free-path spectra and thermoelectric properties in silicon, Europhys. Lett. **109**, 57006 (2015).

[22] Y. Dai, W. Zhou, H.-J. Kim, Q. Song, X. Qian, T.-H. Liu, *et al.*, Simultaneous enhancement in electrical conductivity and Seebeck coefficient by single- to double-valley transition in a Dirac-like band, npj Comput. Mater. **8**, 234 (2022).

[23] P. Giannozzi, O. Andreussi, T. Brumme, O. Bunau, M. B. Nardelli, M. Calandra, *et al.*, Advanced capabilities for materials modelling with Quantum ESPRESSO, J. Phys.: Condens. Matter **29**, 465901 (2017).

[24] J. Y. Zhang, L. J. Zhang, T. Cui, Y. L. Niu, Y. M. Ma, Z. He, *et al.*, A first-principles study of electron–phonon coupling in electron-doped LiH, J. Phys.: Condens. Matter **19**, (2007).

[25] P. A. Varotsos and S. Mourikis, Difference in conductivity between LiD and LiH crystals, Phys. Rev. B **10**, 5220 (1974).

[26] L. Lindsay, Isotope scattering and phonon thermal conductivity in light atom compounds: LiH and LiF, Phys. Rev. B **94**, 174304 (2016).

[27] J. Noffsinger, F. Giustino, B. D. Malone, C.-H. Park, S. G. Louie and M. L. Cohen, EPW: A program for calculating the electron–phonon coupling using maximally localized Wannier functions, Comput. Phys. Commun. **181**, 2140 (2010).

[28] J. Zhou, T.-H. Liu, Q. Song, Q. Xu, Z. Ding, B. Liao, *et al.*, First-principles simulation of electron transport and thermoelectric property of materials, including electron-phonon scattering, defect scattering, and phonon drag, Materials Cloud Archive 2020.106 (2020), doi: 10.24435/materialscloud:5a-7s.

[29] H. Lee, S. Poncé, K. Bushick, S. Hajinazar, J. Lafuente-Bartolome, J. Leveillee, *et al.*, Electron–phonon physics from first principles using the EPW code, npj Comput. Mater. **9**, 156 (2023).

[30] T.-H. Liu, J. Zhou, B. Liao, D. J. Singh and G. Chen, First-principles mode-by-mode analysis for electron-phonon scattering channels and mean free path spectra in GaAs, Phys. Rev. B **95**, 075206 (2017).

[31] G. Brunin, H. P. C. Miranda, M. Giantomassi, M. Royo, M. Stengel, M. J. Verstraete, *et al.*, Electron-phonon beyond Fröhlich: Dynamical quadrupoles in polar and covalent Solids, Phys. Rev. Lett. **125**, 136601 (2020).

[32] V. A. Jhalani, J.-J. Zhou, J. Park, C. E. Dreyer and M. Bernardi, Piezoelectric electron-phonon interaction from ab initio dynamical quadrupoles: Impact on charge transport in wurtzite GaN, Phys. Rev. Lett. **125**, 136602 (2020).

[33] S. Poncé, M. Royo, M. Gibertini, N. Marzari and M. Stengel, Accurate prediction of hall mobilities in two-dimensional materials through gauge-covariant quadrupolar contributions, Phys. Rev. Lett. **130**, 166301 (2023).

[34] J. L. Verble, J. L. Warren and J. L. Yarnell, Lattice dynamics of lithium hydride, Phys. Rev. **168**, 980 (1968).

[35] G. Roma, C. M. Bertoni and S. Baroni, The phonon spectra of LiH and LiD from density-functional perturbation throry, Solid. State. Commun. **98**, 203 (1996).

[36] Z. Xia and H. Zhao, Isotope effects on phonon modes in polyatomic crystals, Phys. Lett. A **473**, 128811 (2023).

[37] G. D. Mahan, Effect of atomic isotopes on phonon modes, Phys. Rev. B **100**, 024307 (2019).

[38] See the Supplemental Material for the electron band structures, mesh convergence, LO coupling strength, isotope mass effect, phonon dispersions, scattering rates, and electron mobilities, which includes Refs. [34, 35, 46, 47].

[39] J. J. Zhou and M. Bernardi, Ab initio electron mobility and polar phonon scattering in GaAs, Phys. Rev. B **94**, 201201 (2016).

[40] J. Ma, Y. Chen and W. Li, Intrinsic phonon-limited charge carrier mobilities in thermoelectric SnSe, Phys. Rev. B **97**, 205207 (2018).

[41] W. Zhou, Y. Dai, T.-H. Liu and R. Yang, Effects of electron-phonon intervalley scattering and band non-parabolicity on electron transport properties of high-temperature phase SnSe: An ab initio study, Mater. Today Phys. **22**, 100592 (2022).

[42] J. J. Zhou, O. Hellman and M. Bernardi, Electron-phonon scattering in the presence of soft modes and electron mobility in $SrTiO_3$ perovskite from first principles, Phys. Rev. Lett. **121**, 226603 (2018).

[43] M. Lundstrom, *Fundamentals of Carrier Transport* (Cambridge University Press, 2009).

[44] Y. Jiang, H. Geng, W. Shi, Q. Peng, X. Zheng and Z. Shuai, Theoretical prediction of isotope effects on charge transport in organic semiconductors, J. Phys. Chem. Lett. **5**, 2267 (2014).

[45] Y. Jiang, H. Geng, W. Li and Z. Shuai, Understanding carrier transport in organic semiconductors: Computation of charge mobility considering quantum nuclear tunneling and delocalization effects, J. Chem. Theory Comput. **15**, 1477 (2019).

[46] S. Poncé, M. Schlipf and F. Giustino, Origin of low carrier mobilities in halide perovskites, ACS Energy Lett. **4**, 456 (2019).

[47] S. Poncé, F. Macheda, E. R. Margine, N. Marzari, N. Bonini and F. Giustino, First-principles predictions of Hall and drift mobilities in semiconductors, Phys. Rev. Research **3**, 043022 (2021).

[48] F. Macheda, P. Barone and F. Mauri, Electron-phonon interaction and longitudinal-transverse phonon splitting in doped semiconductors, Phys. Rev. Lett. **129**, 185902 (2022).

[49] Z. Shuai, W. Li, J. Ren, Y. Jiang and H. Geng, Applying Marcus theory to describe the carrier transports in organic semiconductors: Limitations and beyond, J. Chem. Phys. **153**, 080902 (2020).

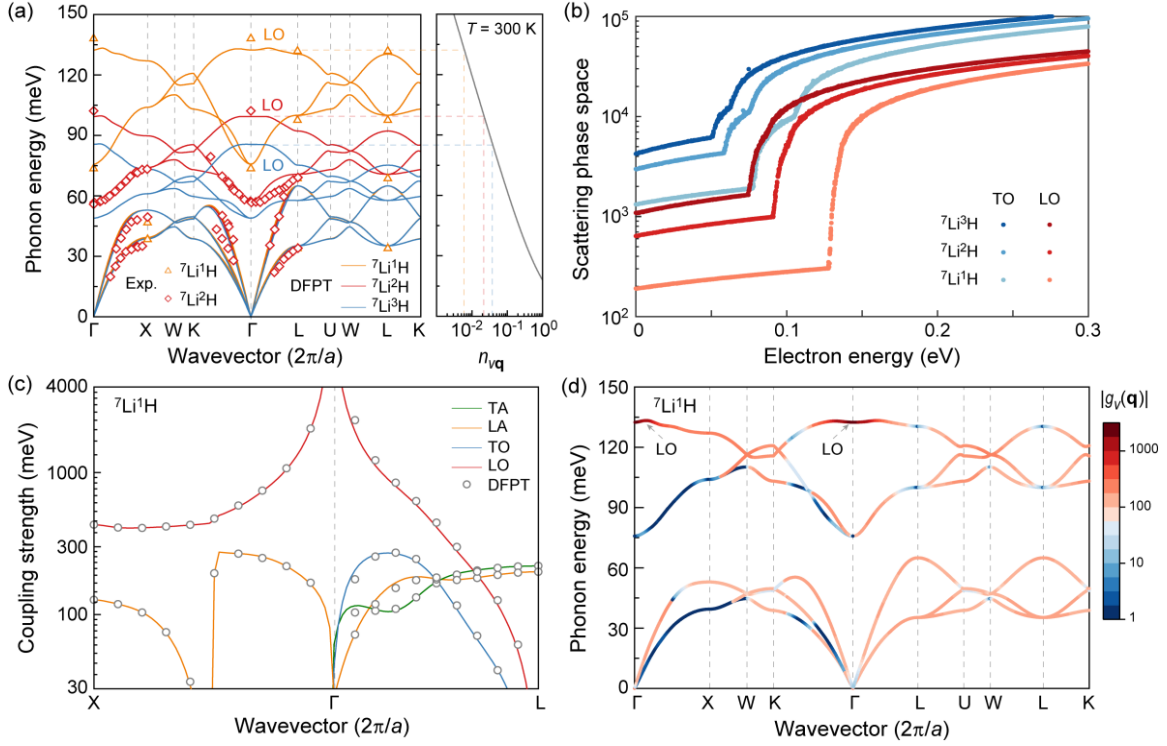

**FIG. 1. Isotope-dependent phonon dispersions and electron-phonon interactions in LiH.** (a) Calculated (solid lines) and measured (symbols) [34, 35] phonon dispersions of $^7$Li$^1$H (yellow), $^7$Li$^2$H (red), and $^7$Li$^3$H (blue). Right panel shows the phonon populations at 300 K. (b) Electron scattering phase space due to the LO (red) and TO (blue) phonons for various isotope compositions. (c) The e-ph coupling strength $|g_\nu(\mathbf{q})| = \sqrt{\sum_{mn} |g_{mn\nu}(\mathbf{k},\mathbf{q})|^2/N}$ for $^7$Li$^1$H obtained using Wannier interpolation (solid lines) and via direct DFPT calculations (gray circles), with the CBM as the initial electron state and summation over the lowest conduction band. (d) Mode-resolved coupling strength overlaid on top of the phonon dispersion of $^7$Li$^1$H using a log-scale colormap.

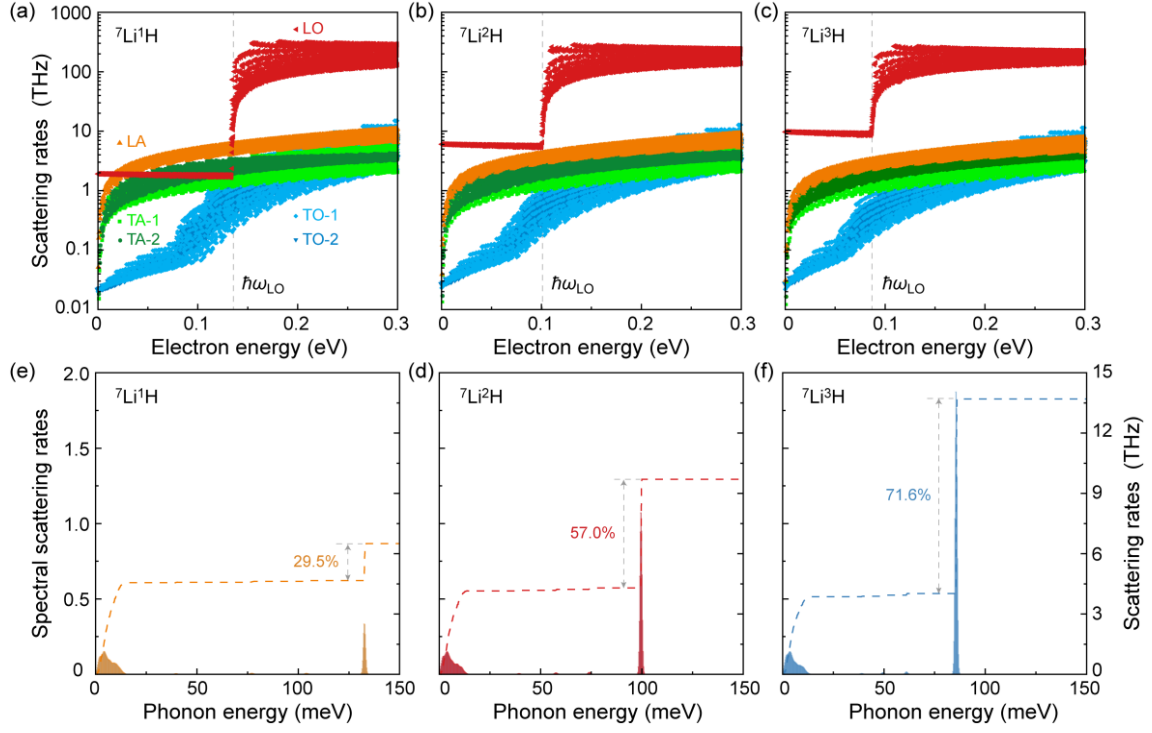

**FIG. 2. Variation of mode-resolved electron-phonon scattering rates with isotope compositions dominated by the Fröhlich interaction.** (a) $^7$Li$^1$H, (b) $^7$Li$^2$H, and (c) $^7$Li$^3$H. Phonon spectral decomposition (shaded areas, left axes, THz/meV) and cumulative scattering rates (dashed lines, right axes) for (d) $^7$Li$^1$H, (e) $^7$Li$^2$H, and (f) $^7$Li$^3$H, respectively. The electron states with an average energy of 39 meV away from the conduction band edge are selected [46, 47]. Contributions from the Fröhlich interaction are marked with gray arrows.

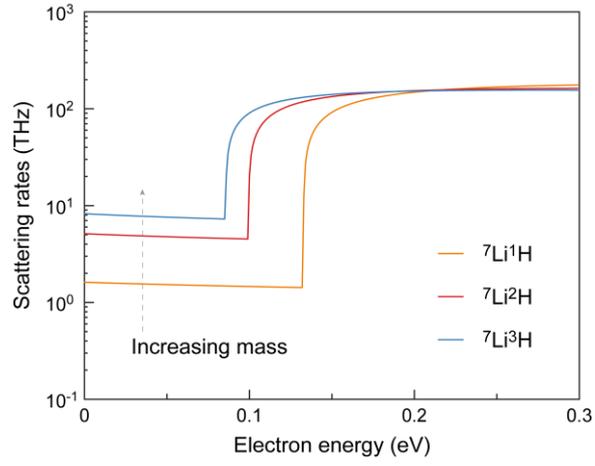

**FIG. 3. Isotope mass effect on Fröhlich scattering based on the phenomenological model.**

Key parameters in Eq. (3) include $m_e = 0.36\ m_0$, $T = 300$ K, $\epsilon_\infty = 5.14$, and $\epsilon_0 = 12$.

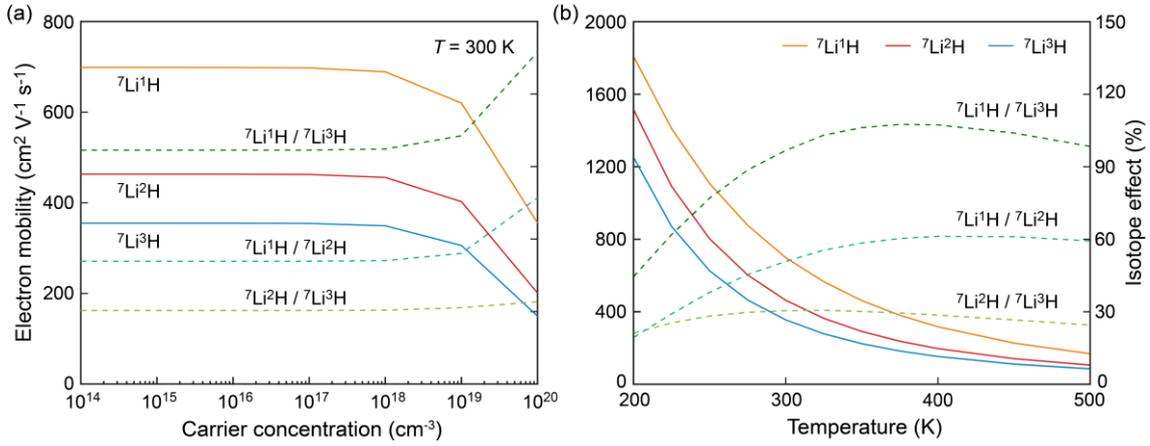

**FIG. 4. Isotope effect on electron mobility.** Calculated electron mobilities (solid lines) of $^7$Li$^1$H (yellow), $^7$Li$^2$H (red), and $^7$Li$^3$H (blue) as a function of (a) carrier concentration, and (b) temperature. The dashed green lines in both panels indicate the mobility enhancement (right axes) upon substitution of the hydrogen isotopes as labeled. Power-law fittings of the temperature-dependent electron mobility in (b) yield $T^{-2.62}$, $T^{-2.94}$, and $T^{-2.96}$ for $^7$Li$^1$H, $^7$Li$^2$H, and $^7$Li$^3$H, respectively. Details of the fitting are shown in Fig. S7 [38]. A carrier concentration of $10^{12}$ cm$^{-3}$ is assumed for (b), although (a) starts at $10^{14}$ cm$^{-3}$ to highlight the trend at higher concentrations.